\documentclass[aps,amsmath,amssymb,superscriptaddress,prl,10pt,twocolumn,eqrefnum,nofootinbib]{revtex4-2}

\usepackage{graphicx}
\usepackage{dcolumn}
\usepackage{bm}
\usepackage{physics}
\usepackage{color,xcolor}
\usepackage[hyperfootnotes=false]{hyperref}
\usepackage[normalem]{ulem}

\usepackage{lipsum}

\newcommand{\upup}{{\uparrow\uparrow}}
\newcommand{\dodo}{{\downarrow\downarrow}}

\newcommand{\cc}{\mathrm{ch}}
\newcommand{\sn}{\mathrm{sp}}

\begin{document}

\preprint{APS/123-QED}

\title{Quantum-geometric spin and charge Josephson diode effects}

\author{Niklas L. Schulz}
\email{niklas.schulz@uni-greifswald.de}
\author{Danilo Nikoli\'c}%
\email{danilo.nikolic@uni-greifswald.de}
\author{Matthias Eschrig}
\email{matthias.eschrig@uni-greifswald.de}
\affiliation{Institute of Physics, University of Greifswald, Felix-Hausdorff-Strasse 6, 17489 Greifswald, Germany}

\date{\today}
\begin{abstract}
We present a general mechanism for large charge and spin Josephson diode effects in strongly spin-polarized superconductor-ferromagnet hybrid structures with a noncoplanar spin texture, formulated in terms of quantum-geometric phases. We present necessary conditions for this effect to occur, and show numerical results for disordered materials, relevant for applications. We calculate Josephson diode efficiencies for both charge and spin diodes and show that a spin diode efficiency of 100\% can be reached. Finally, we present a SQUID device that can switch between nearly pure spin-up and spin-down equal-spin supercurrents across the ferromagnet by reversing the flux. These findings establish functionalities that are absent for coplanar spin textures.
\end{abstract}

\maketitle


Superconducting spintronics has attracted increasing interest both from the fundamental point of view of realizing new states of matter and from the practical point of view as potential contributor to an energy-saving technology for large-scale computing and data storage centers in an era where energy consumption has become a critical issue \cite{eschrigSpinpolarizedSupercurrentsSpintronics2015,linderSuperconductingSpintronics2015,yangBoostingSpintronicsSuperconductivity2021,caiSuperconductorFerromagnetHeterostructures2023}. The production and control of spin-polarized supercurrents is essential for superconducting spintronics. Whereas at present, spin-triplet supercurrents can be produced routinely \cite{keizerSpinTripletSupercurrent2006,khaireObservationSpinTripletSuperconductivity2010,anwarLongrangeSupercurrentsHalfmetallic2010,robinsonControlledInjectionSpinTriplet2010,birgereview2018,Glick2018,Caruso2019,Aguilar2020,birge2024}, the production and control of pure spin supercurrents as well as new functionalities of devices like diode effects are still open questions. Especially promising devices are those containing strongly spin-polarized ferromagnetic materials, across which phase-coherence is maintained only within each spin band, however not between the two spin bands \cite{greinSpinDependentCooperPair2009,eschrigSpinpolarizedSupercurrentsSpintronics2011,eschrigSpinpolarizedSupercurrentsSpintronics2015}. In this case, the Josephson phases in the two spin bands decouple, leading to a new channel of control via spin-geometric phases that directly add to the Josephson phases with opposite sign in the two spin bands \cite{greinSpinDependentCooperPair2009}. As a result, as we will show in this article, new phenomena occur that are entirely governed by  quantum-geometric phases that appear due to noncoplanar spin textures in the device.

\begin{figure}[b!]
    \centering
    \includegraphics[width=\linewidth]{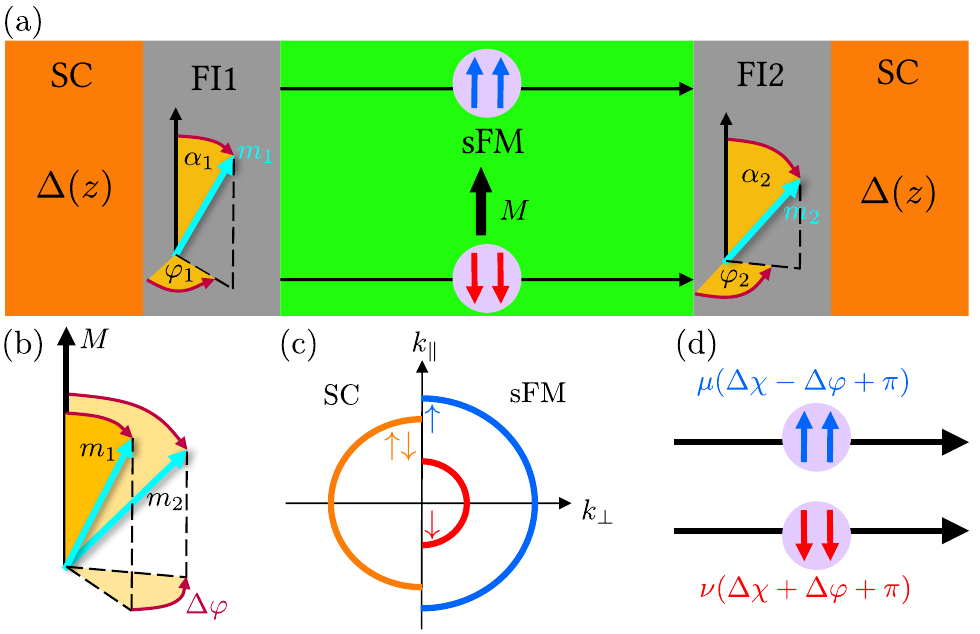}
    \caption{(a) Superconducting hybrid structure consisting of a strongly spin-polarized metallic ferromagnet (sFM) coupled to singlet superconductors (SC) by thin ferromagnetic insulating layers (FI1/2). At the outer interfaces the system is connected to superconducting reservoirs each characterized by $\Delta_{1/2} e^{i\chi_{1/2}}$(Left: 1, Right: 2). (b) Relative orientation of the FI magnetic moments $m_{1/2}$ with respect to the sFM magnetization $M$, defining a quantum-geometric phase $\Delta \varphi$. (c) Fermi surface mismatch between SC and sFM, shown for the case $k_{F,\downarrow}^\mathrm{sFM} < k_F^{\mathrm{SC}} < k_{F,\uparrow}^\mathrm{sFM}$. (d) Sketch of phases acquired by the transmission of $\mu $ $\upup$-pairs and $\nu $ $\dodo$-pairs.$^1$}
    \label{fig:Plots_system_under_study}
\end{figure}

Quantum-geometric phases play an important role in materials with nontrivial spin-structure like altermagnets \cite{Altermagnetism_review2}, topological insulators \cite{bernevigQuantumSpinHall2006,hsiehTopologicalDiracInsulator2008,andoTopologicalInsulatorMaterials2013}, or skyrmionic materials \cite{rosslerSpontaneousSkyrmionGround2006,yuRealspaceObservationTwodimensional2010,Magnetic_skyrmions2017}. In general, a noncoplanar spin arrangement is necessary for nontrivial effects based on quantum-geometric phases to appear. In superconductor-ferromagnet hybrid structures, such as shown in Fig.~\ref{fig:Plots_system_under_study} (a), spin-structure plays an multi-facetted role. 
\footnotetext[1]{The vectors $M$, $m_1$, and $m_2$ are here defined in units of the quasiparticle magnetic moment (which can be negative for spin-up as, e.g. in Fe, Co, Ni), i.e. they point in the directions of the respective exchange fields.}

Spin-rotational invariance is broken at interfaces between a spin-singlet superconductor and a ferromagnet, leading to the appearance of spin-triplet pair correlations near the interface. These are short-ranged in the ferromagnet. In order to transform them into long-ranged, equal-spin pair correlations inside the ferromagnet, a noncollinear spin profile is necessary. This leads to the mechanism of spin-mixing and triplet rotation for the creation of long-range triplet supercurrents \cite{eschrigSpinpolarizedSupercurrentsSpintronics2011}. Novel phenomena occur, when the spin-profile in the hybrid structure is not only noncollinear, but noncoplanar, as shown in Fig.~\ref{fig:Plots_system_under_study} (b), opening up the field to an entire new set of phenomena related to the interplay between quantum-geometric phases and Josephson phases.
In this case, a nonzero solid angle spanned by three mutually noncoplanar magnetization vectors leads to a nontrivial quantum-geometric phase in the equal-spin pair correlation function, see Fig.~\ref{fig:Plots_system_under_study} (d) \cite{greinSpinDependentCooperPair2009,eschrigPhasesensitiveInterfaceProximity2019}. In this article, we show, that this quantum-geometric phase in combination with ferromagnetic spin-filtering leads to the appearance of a Josephson diode effect (JDE) \cite{huProposedDesignJosephson2007a,Reynoso2008,greinSpinDependentCooperPair2009,Margaris_2010,ilicTheorySupercurrentDiode2022,soutoJosephsonDiodeEffect2022,davydovaUniversalJosephsonDiode2022a,hePhenomenologicalTheorySuperconductor2022,Karabassov2022,nadeemSuperconductingDiodeEffect2023,Meyer2024,sunGatetunableSignReversal2024,andoObservationSuperconductingDiode2020, baumgartnerSupercurrentRectificationMagnetochiral2022,strambiniSuperconductingSpintronicTunnel2022, houUbiquitousSuperconductingDiode2023, trahmsDiodeEffectJosephson2023} in both the spin- and charge channel (in a JDE, the critical Josephson currents for positive and negative current directions differ). Our mechanism is non-local and does neither involve finite-momentum pairing nor spin-orbit interaction; therefore, it fundamentally differs in these respects from other models cited above, and builds on ideas outlined in Ref.~\cite{greinSpinDependentCooperPair2009}. We also show that for given spin-texture there is a certain Josephson phase resulting in a 100\% spin-polarized supercurrent across the ferromagnet; the spin polarization can be reversed by reversing the Josephson phase across the junction (e.g., by reversing the magnetic flux in a SQUID geometry). Thus, we propose a device that can switch between spin-up and spin-down supercurrents with nearly 100\% spin-polarization.

We use the quasiclassical Green's function method suitable for treating diffusive systems~\cite{belzigQuasiclassicalGreensFunction1999}. 
We determine numerically the Josephson current-phase relation (CPR) in the system shown in Fig.~\ref{fig:Plots_system_under_study}(a), consisting of two superconductors (SC) adjacent to a strongly spin-polarized metallic ferromagnet (sFM). Our model includes a thin ferromagnetic insulating layer on either side of the metallic ferromagnet. The superconductors are connected to superconducting reservoirs at the outer interfaces ensuring a well-defined phase bias across the junction, $\Delta\chi=\chi_2-\chi_1$. The  Green's function is a 4$\times$4 matrix in combined particle-hole $\otimes$ spin space obeying the Usadel transport equation \cite{usadelGeneralizedDiffusionEquation1970} (we set $\hbar=k_B=1$)
\begin{equation}
    \comm{E \hat{\tau}_3 - \hat{\Delta}}{\hat{G}} - iD \bm{\nabla} (\hat{G}\bm{\nabla}\hat{G}) = \hat{0}, \label{eq:Usadel}
\end{equation}
with the normalization condition $\hat{G}^2 = \hat{1}$. Here, $\hat{\dots}$ denotes the
matrix structure, $\hat{\tau}_3$ is the third Pauli matrix in particle-hole space, $E$ denotes the quasiparticle energy, and $D=v_F\ell/3$ is the diffusion coefficient with $v_F$ being the Fermi velocity and $\ell=v_F\tau$ the elastic mean free path. Finally, the BCS self-energy is given by the $\hat{\Delta}$ term which is treated self-consistently
\begin{equation}
    -\Delta(\Vec{r})\ln\frac{T}{T_c} = \int\limits_{-\infty}^\infty \frac{dE}{2} \left[\mathcal{F}_0(E,\Vec{r})+\frac{\Delta(\Vec{r})}{E}\right]\tanh\frac{E}{2T},
\end{equation}
where $\mathcal{F}_0$ denotes the spin-singlet pairing amplitude. The relevant length scale for the junction is set by the superconducting coherence length $\xi = \sqrt{D / T_c}$, where $T_c$ is the critical temperature of the superconductor. Accordingly, the diffusive limit is achieved by the condition $\ell\ll\xi$. 

We consider a strongly spin-polarized metallic ferromagnet, for which each spin-band must be treated separately with its own Fermi velocity $\Vec{v}_{F\eta}$ and Fermi momentum $\Vec{p}_{F\eta}$ ($\eta=\uparrow,\downarrow$) \cite{greinSpinDependentCooperPair2009}.
Consequently, these have effectively different diffusion coefficients $D_\eta$ that enter the Usadel equation~\eqref{eq:Usadel}. In this approach, the coherence across the junction is maintained only for equal-spin pair correlations~\cite{schulz2025PRB}, therefore, the Green's functions in the metallic ferromagnet are spin scalars denoted as $\breve{G}_{\eta\eta}$ opposing to the Green's functions in the superconductors which are $2\times 2$ spin matrices. As we show below, the FI layers induce a nontrivial coupling between the spin bands in the metallic ferromagnet manifesting itself through the anomalous CPR exhibiting the JDE, and which is calculated from
\begin{equation}
\label{eqn:current}
    I_{\eta\eta} = N_{F \eta} D_\eta \Re\! \int\limits_{-\infty}^\infty \frac{dE}{8} \mathrm{Tr}\left\{\breve{\tau}_3 \breve{G}_{\eta\eta}\bm{\nabla} \breve{G}_{\eta\eta}\right\} \tanh{\frac{E}{2T}}.
\end{equation}

The ferromagnetic insulating (FI) layers located at the SC/sFM interfaces are assumed to be of a thickness and a strength comparable to the Fermi wavelength $\lambda_F$  and the Fermi energy $E_F$, respectively, and thereby cannot be described by the quasiclassical theory. Instead, they enter the model as  effective boundary conditions for the Usadel equation~\eqref{eq:Usadel}. For implementing them, we follow Ref.~\cite{eschrigGeneralBoundaryConditions2015} characterizing the interface region by a normal-state scattering matrix obtained by matching Bloch waves from the superconductor and the metallic ferromagnet. For definiteness, we use a Fermi surface geometry as shown in Fig.~\ref{fig:Plots_system_under_study} (c).
We model the scattering potential as a spin-split box potential following Refs.~\cite{greinSpinDependentCooperPair2009,greinTheorySuperconductorferromagnetPointcontact2010} such that it is characterized by spin-dependent bias potentials $V_{\uparrow}=1.1 E_F$ and $V_{\downarrow}=1.9 E_F$, the exchange splitting $J=0.8E_F$, and the width of the scattering region $d=0.6\lambda_F/2\pi$. For definiteness, we assume identical FI materials with the magnitudes of the exchange splitting equal to that in the the metallic ferromagnet. The phenomena we wish to discuss are not sensitive to the precise magnitude of the FI magnetic moments.
The magnetic moment of each FI layer points in a direction characterized by the spherical angles $\alpha$ and $\varphi$ taken with respect to magnetization in the metallic ferromanget which sets the global quantization axis [see Fig.~\ref{fig:Plots_system_under_study}(a)].\footnotemark[1] This parameterization gives rise to the relative azimuthal angle across the junction, $\Delta\varphi=\varphi_2-\varphi_1$ [see Fig. \ref{fig:Plots_system_under_study}(b)], which, as we discuss below, enters the CPR as a  \textit{quantum-geometric phase} and which turns out to be one of the crucial ingredients for the appearance of JDE in our system. Besides this, a nonzero value of the polar angle $\alpha_i$ at at least one of the two FI layers is required for the occurrence of the triplet-rotation mechanism necessary for the appearance of long-range spin-triplet supercurrents across the ferromagnet, without which a Josephson effect would not occur in strongly spin-polarized junctions. In our calculations, we take $\alpha_1=\alpha_2=\pi/2$. In addition, the thicknesses of the superconductors and the metallic ferromagnet are taken to be $L_\mathrm{SC}=5\xi$ and $L_\mathrm{sFM}=\xi$, respectively, and the temperature is $T=0.1T_c$. 

The spin geometric phase $\Delta \varphi$ is of similar fundamental importance to the junction as the Josephson phase $\Delta \chi =\chi_2-\chi_1$ resulting from the superconducting pair potentials $\Delta_1=|\Delta_1|e^{i\chi_1}$ and $\Delta_2=|\Delta_2|e^{i\chi_2}$ on either side of the  junction \cite{eschrigPhasesensitiveInterfaceProximity2019}. It directly enters the CPR for each spin band with opposite sign for the two spin projections \cite{greinSpinDependentCooperPair2009}. The transmission of each $\uparrow\uparrow$-pair involves a Josephson phase of $\Delta\chi-\Delta \varphi + \pi$, and the transmission of each $\downarrow\downarrow$-pair involves a Josephson phase of $\Delta\chi+\Delta \varphi + \pi$ [see Fig.~\ref{fig:Plots_system_under_study}(d)]. Consequently, the coherent transmission event of $\mu$ $\uparrow\uparrow$-pairs and $\nu$ $\downarrow\downarrow$-pairs involves a total phase of $(\mu+\nu)\Delta \chi - (\mu-\nu) \Delta \varphi + (\mu+\nu)\pi $. The CPRs for each spin band can then be obtained as a sum over all coherent $(\mu,\nu)$-multiple-pair transmission events in either direction (positive or negative $\mu$ or $\nu$) with the corresponding Josephson phase. As a result, together with the condition
$I_{\uparrow\uparrow (\downarrow\downarrow)}(\Delta\chi,\Delta\varphi) = - I_{\uparrow\uparrow (\downarrow\downarrow)}(-\Delta\chi,-\Delta\varphi)$ resulting from an analysis of the system under time reversal, the spin-resolved Josephson current can be written in a Fourier expansion as
\begin{align}
    I_{\uparrow\uparrow} (\Delta \chi, \Delta \varphi) &= \frac{1}{2}\sum_{\mu,\nu=-\infty}^\infty \mu (-1)^{\mu+\nu} I_{\mu,\nu} \sin\psi_{\mu,\nu}, 
    \label{Iuu} \\
    I_{\downarrow\downarrow} (\Delta \chi, \Delta \varphi) &= \frac{1}{2}\sum_{\mu,\nu=-\infty}^\infty \nu (-1)^{\mu+\nu} I_{\mu,\nu} \sin\psi_{\mu,\nu},
    \label{Idd}
\end{align}
with $I_{-\mu,-\nu}=I_{\mu,\nu}$ and the \textit{effective Josephson phase}
\begin{align}
    \psi_{\mu,\nu}(\Delta \chi, \Delta \varphi) &=(\mu+\nu)\Delta\chi - (\mu-\nu)\Delta\varphi .
    \label{eJp}
\end{align}
Our numerical calculations of the CPR using Green's functions, presented below, not only fully confirm this intuitively appealing picture, which was introduced in Refs. \cite{eschrigSpinpolarizedSupercurrentsSpintronics2015,eschrigPhasesensitiveInterfaceProximity2019} but also extend it to the case of diffusive systems. For details we refer to our accompanying paper \cite{schulz2025PRB}. We note that Eqs.~\eqref{Iuu}-\eqref{eJp} imply that the charge supercurrent $I_{\rm charge}$ and spin supercurrent $I_{\rm spin}$ can be obtained from the Josephson energy $E_J$ via
\begin{align}
    E_J(\Delta \chi, \Delta \varphi) =-\frac{\hbar}{2}& \sum_{\mu,\nu=-\infty}^\infty (-1)^{\mu+\nu}I_{\mu,\nu}\cos \psi_{\mu,\nu},
    \label{JE}\\
    I_{\rm charge}(\Delta \chi, \Delta \varphi)&=2e(I_{\uparrow\uparrow}+I_{\downarrow\downarrow})=
    \frac{2e}{\hbar}\frac{\partial E_J}{\partial \Delta \chi},\\
    I_{\rm spin }(\Delta \chi, \Delta \varphi)&=2S (I_{\uparrow\uparrow}-I_{\downarrow\downarrow})= -\frac{2S}{\hbar}\frac{\partial E_J}{\partial \Delta \varphi}.
\end{align}
with $S=\frac{\hbar}{2}$, and $e<0$ denotes the electron charge.

The quantities $I_\cc=I_{\uparrow\uparrow}+I_{\downarrow\downarrow}$, and $I_\sn=I_{\uparrow\uparrow}-I_{\downarrow\downarrow}$ can then be discussed in terms of the Fourier components of the CPR.
Critical Josephson currents in positive ($+$) and negative ($-$) directions are defined as $I_\cc^+=\max_{\Delta\chi} (I_\cc )$ and $I_\cc^-=\min_{\Delta \chi} (I_\cc )$, and the corresponding Josephson phases are $\Delta\chi^+=\mbox{argmax}_{\Delta\chi} (I_\cc )$ and $\Delta\chi^-=\mbox{argmin}_{\Delta\chi} (I_\cc )$.
The charge JDE occurs when $|I_\cc^+|\ne |I_\cc^-|$. We define diode efficiencies for ${\rm x}=\cc, \sn$ as ${\eta_{\rm x} =(|I_{\rm x}^+|-|I_{\rm x}^-|)/(|I_{\rm x}^+|+|I_{\rm x}^-|)}$, where $I^\pm_{\rm x} = I_{\rm x} (\Delta\chi^\pm)$.

Our numerical Fourier analysis \cite{schulz2025PRB} shows that the CPR is closely reproduced by the terms with $|\mu |+|\nu |\le 2$ in Eqs.~\eqref{Iuu}-\eqref{Idd}. 
Restricting to these terms and analysing the relative magnitudes of the Fourier components of the CPR, we find that the maximal magnitude $|\widehat \eta_\cc |$ of the  charge Josephson diode efficiency is obtained as 
\begin{align}
    |\widehat\eta_\cc |\approx  \frac{4\sqrt{2}|AB|-B^2}{(\sqrt{8}|A|+|B|)^2},\quad |B| \le \sqrt{8}|A|
    \label{etamodel1}
\end{align}
with $A=2(I_{1,1}-I_{2,0}-I_{0,2})$, $B=I_{1,0}-I_{0,1}$, and $|\widehat\eta_\cc |\lesssim |A|/(\sqrt{2}|B|)$ for $|B| > \sqrt{8}|A|$.
 The maximal efficiency that can be obtained from Eq.~\eqref{etamodel1} is for $|B|=\sqrt{2}|A|$ and is 33.$\overline{3}$\%. 
The JDE is absent when either $A$ or $B$ vanish. We find that $I_{1,1}$ dominates in $A$ and is crucial for a large effect to occur. If $A$ is zero,
then the CPR can be expressed as a shifted sine function, leading to a ``$\phi_0$-junction'' \cite{GeshkenbeinLarkin1986,Sigrist1998,BraudeNazarov2007,Asano2007,eschrigTripletSupercurrentsClean2008,Buzdin2008,BeenakkerBrouwer2009} and an anomalous Josephson effect, but no JDE. A nonzero $B$ requires $I_{1,0}\ne I_{0,1}$. The JDE is also absent for $\Delta \varphi = k\frac{\pi}{2}$, $k$ integer, as then Eqs.~\eqref{Iuu}-\eqref{eJp} lead to CPRs that are either anti-symmetric around $\Delta\chi=k\pi $ or $\Delta\chi=\frac{\pi}{2}+k\pi$.
Thus, a nonzero JDE in the charge channel occurs under the following conditions: (a) the geometric solid angle spanned by the three magnetic vectors in Fig.~\ref{fig:Plots_system_under_study} is nonzero; (b) $\Delta \varphi \ne k\frac{\pi}{2}$; (c) the densities of states of the two spin-bands in the metallic ferromagnet differ, i.e., $I_{0,1}\ne I_{1,0}$; and (d) the $I_{1,1}$ (crossed pair transmission \cite{greinSpinDependentCooperPair2009}) terms
are appreciable. The last condition requires a ferromagnet with two spin bands (no half-metal), the third condition requires a strongly spin-polarized ferromaget, the second condition is a geometric restriction, and the first conditions requires a noncoplanar spin texture. In the following, we will present our fully self-consistent numerical results confirming this appealing picture and showing that indeed large Josephson-diode effects appear. 

\begin{figure}[t!]
    \centering
    \includegraphics[width=\linewidth]{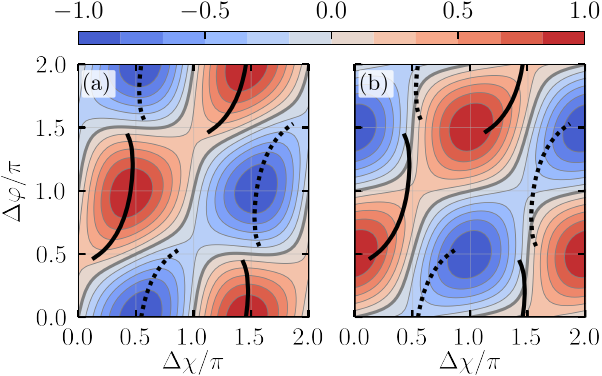}
    \caption{The functional dependence of (a) the charge current $I_\cc$ and (b) the spin current $I_\sn$ on the quantum-geometric phase difference $\Delta\varphi = \varphi_2 - \varphi_1$ and the superconducting phase difference $\Delta\chi = \chi_2 - \chi_1$ normalised to their overall maximum value reached. The thick grey lines denote zero current. In both panels the black solid and dashed lines show $\Delta\chi^+(\Delta\varphi)$ and $\Delta\chi^-(\Delta\varphi)$ for the positive and negative critical charge current at given $\Delta\varphi$, respectively.}
    \label{fig:colormap_all_curr}
\end{figure}

Whereas $I_{1,1}$ enters only the charge current, a term $-I_{1,-1}\sin(2\Delta\varphi)$ is only present in the spin current and is $\Delta\chi$-independent. It follows that positive and negative extremal spin currents differ as soon as $I_{1,-1}$ is nonzero and $\Delta\varphi \ne k\frac{\pi}{2}$.
This is a generic case for our setup, and for strongly asymmetric junctions the $I_{1,-1}$ term even dominates. As this term is decoupled from the Josephson phase $\Delta\chi$, we define the spin diode efficiency $\eta_\sn $ for the spin Josephson currents via the charge Josephson phases $\Delta \chi^\pm$, rather than for the maximal and minimal spin Josephson currents. It still
can reach 100\% when the spin current vanishes at either $\Delta\chi^+$ or $\Delta\chi^-$ for a given $\Delta\varphi =\widehat{\Delta \varphi}_\sn$. 
In symmetric junctions, where $I_{1,-1}$ is rather small, we obtain $|\widehat\eta_\sn| =1$ for 
\begin{align}
    \widehat{\Delta \varphi}_\sn \approx \pm \arcsin\left( \frac{2B}{4(A+C)-CB^2/A^2}\right)
    \label{etasp1}
\end{align}
with $C=2(I_{1,-1}+I_{2,0}+I_{0,2})$, for $|B|<2|A|$, $|C|\ll |A|$.

In Fig.~\ref{fig:colormap_all_curr}, we show the self-consistently calculated [see Eq.~\eqref{eqn:current}] charge [panel (a)]
and spin [panel (b)] currents as functions of $\Delta\chi$ and $\Delta\varphi$ for the junction shown in Fig.~\ref{fig:Plots_system_under_study}(a). As mentioned above, the junction is symmetric apart from the relative misalignment angle $\Delta \varphi $ of the magnetic moments of the FI layers. The CPRs of both the charge and the spin current are strongly varying with $\Delta\varphi$. The black lines denote the Josephson phases corresponding to the critical charge current in the positive ($\Delta\chi^+$; solid) and the negative ($\Delta\chi^-$; dotted) direction introduced above. 

\begin{figure}
    \centering
    \includegraphics[width=\linewidth]{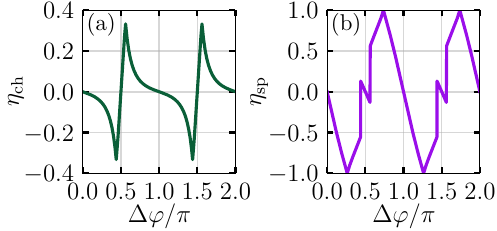}
    \caption{The functional dependence of (a) the charge diode efficiency $\eta_\cc$ (green) and (b) the spin diode efficiency $\eta_\sn$ (violet) on the quantum-geometric phase difference $\Delta\varphi$ for the currents shown in Fig.~\ref{fig:colormap_all_curr}.}
    \label{fig:selected_diode_eff_comparison}
\end{figure}

In Fig.~\ref{fig:selected_diode_eff_comparison} 
the charge diode efficiency [panel (a)] and the spin diode efficiency [panel (b)] are shown. The spin JDE is observed by tuning the junction to its critical charge currents, therefore it is determined by the same Josephson phases $\Delta\chi^\pm$ as the charge JDE \cite{sunGatetunableSignReversal2024}.
 Since the CPR, and therefore the critical currents, are functions of the quantum-geometric phase difference $\Delta\varphi$ the diode efficiency is also a function of $\Delta\varphi$. 
The maximal charge diode efficiency for the symmetric junction case is reached at $\eta_\cc^\mathrm{max} \approx 33.3 \%$,
which compares well to other theoretical predictions \cite{davydovaUniversalJosephsonDiode2022a,soutoJosephsonDiodeEffect2022,ilicTheorySupercurrentDiode2022,Meyer2024}. To compare with Eq.~\eqref{etamodel1}, we compute the leading Fourier coefficients from the numeric CPRs, $(I_{0,1},I_{1,1},I_{1,-1},I_{2,0},I_{0,2})/I_{1,0}\approx (0.680,0.084, 0.027, -0.015, -0.008)$, and
obtain $\abs{\widehat\eta_\cc} \approx 33.3 \%$ 
which is in excellent agreement with the numerical value. 
The spin diode efficiency reaches maximal values of 100\%. These occur when for given $\Delta\varphi$ the spin current is zero for $\Delta\chi^+$ and nonzero for $\Delta\chi^-$ or
the other way around, see Fig.~\ref{fig:colormap_all_curr}(b). Equation~\eqref{etasp1} yields with the above Fourier coefficients $\widehat{\Delta\varphi}_\sn \approx \pm 0.26\pi +k\pi$, 
with $k$ integer, in excellent agreement with 
Fig.~\ref{fig:selected_diode_eff_comparison}(b).
For an extensive discussion of the charge- and spin-diode efficiencies see our accompanying paper~\cite{schulz2025PRB}.

We draw particular attention to the contributions with $\mu=\nu$.
The simultaneous presence of an $\uparrow\uparrow$-pair and a $\downarrow\downarrow$-pair in the metallic ferromagnet as well as two spin-singlet pairs in the superconductor can give rise to a transfer of two Cooper pairs across the junction  without the need of the triplet-rotation mechanism at one of the interfaces. Thus, processes accompanied by the transfer of the same number of $\uparrow\uparrow $-pairs and $\downarrow\downarrow$-pairs, called {\it crossed pair transmission} processes \cite{greinSpinDependentCooperPair2009}, are enhanced compared to other processes, which require the triplet rotation mechanism to act on both interfaces for all surplus pairs in one spin band compared to the other. 

We also point out that in a half-metal the JDE is absent. This follows from Eqs.~\eqref{Iuu}-\eqref{eJp}, as $\nu=0$ leads to $\psi_{\mu,0}=\mu \cdot (\Delta\chi-\Delta\varphi)$, and a $\phi_0$-junction with $\phi_0=\Delta\varphi$ is obtained (the CPR is antisymmetric around $\phi_0$).
\begin{figure}
    \centering
    \includegraphics[width=0.85\linewidth]{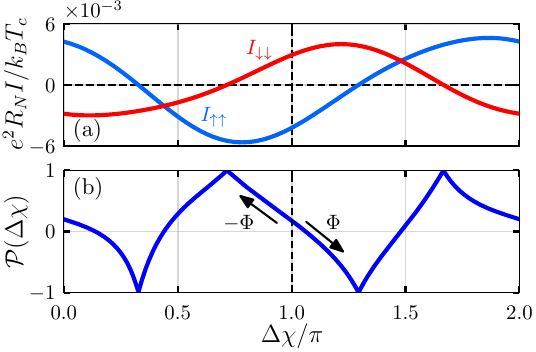}
    \caption{(a) The spin-resolved currents in the metallic ferromagnet as functions of $\Delta\chi$ [see Fig.~\ref{fig:colormap_all_curr}] for $\Delta\varphi = 0.31 \pi$. (b) The corresponding spin-polarization $\mathcal{P}(\Delta\chi)$ for the Josephson currents shown in (a). $\Phi$ denotes the flux through a SQUID geometry and the arrows denote the change of the current in response to a nonzero flux.}
    \label{fig:polarisation}
\end{figure}

We finally propose a switchable device to produce a fully spin-polarized supercurrent. In Fig.~\ref{fig:polarisation}(a) we show for a symmetric junction the spin up and spin-down supercurrents as a function of $\Delta \chi $ for given $\Delta \varphi =0.31 \pi $. We define the spin-polarization of the charge current as $\mathcal{P}(\Delta\chi) = (|I_\upup| - |I_\dodo|) / (|I_\upup| + |I_\dodo|)$ and show it in Fig.~\ref{fig:polarisation}(b). It is possible to adjust $\Delta \chi $ such that the current is zero in one of the two spin bands and nonzero in the other, leading to a 100\% spin polarized current. This can be achieved, e.g., in a SQUID geometry by adjusting the flux through the loop.  Furthermore, two such realizations of fully spin-polarized Josephson currents correspond approximately to reversed fluxes of equal magnitude through the loop. By relaxing the condition that the current vanishes exactly in one of the bands, one can switch between two nearly fully spin-polarized supercurrents ($\mathcal{P} \approx \pm 1$), polarized in opposite directions, by reversing the direction of the flux through the loop [see Fig.~\ref{fig:polarisation}(b)]. This functionality is another example for a new paradigm in the field of superconducting spintronics resulting from noncoplanar spin textures in strongly spin-polarized Josephson devices and the physics associated with crossed pair transmission processes and supercurrent spin-filtering.

In conclusion, we have presented a study of the spin and charge Josephson diode effect as well as the spin switching effect occurring in strongly spin-polarized ferromagnetic trilayers. We have shown that the crucial role in both effects is played by the quantum-geometric phase determined by the relative azimuthal angle between the magnetizations of the ferromagnetic FI layers. Our numerical calculations predict a large charge JDE with efficiency up to $\eta_\cc \approx 33\%$ and a spin JDE that can reach perfect efficiency, $\eta_\sn=100\%$. Finally, we propose a way of switching between nearly pure spin-up and spin-down supercurrents in a SQUID geometry involving the discussed junction. Our theory can be experimentally tested in Josephson junctions involving strongly polarized ferromagnetic trilayers based on, e.g., metallic Ni or Co compounds~\cite{khaireObservationSpinTripletSuperconductivity2010,Glick2018,Aguilar2020} and  ferromagnetic  insulators  GdN~\cite{Caruso2019} or EuS/EuO~\cite{Moodera2007}. The spin current in the central ferromagnet can be measured by appropriate experimental techniques, e.g., via proximity-induced inverse superspin Hall effect~\cite{Linder2019}. Incorporated in a SQUID this junction can exhibit spontaneous fluxes $<\Phi_0/2$ as a signature of the anomalous Josephson effect.

We thank Ralf Schneider for fruitful discussions. NLS and ME acknowledge funding by the Deutsche Forschungsgemeinschaft (DFG, German Research
Foundation) under project number 530670387. The computations were enabled by resources provided by the University Computer Centre of the University of Greifswald.

The data that support the findings of this article are openly available \cite{dataset}.

\end{document}